\documentclass[letterpaper,twocolumn,prl,aps,superscriptaddress,amsmath,amssymb,floatfix]{revtex4-2}
\usepackage{mathptmx}
\usepackage[latin9]{luainputenc}
\usepackage{color}
\usepackage{amsmath}
\usepackage{amssymb}
\usepackage{graphicx}
\usepackage{esint}
\usepackage[unicode=true,
 bookmarks=true,bookmarksnumbered=false,bookmarksopen=false,
 breaklinks=false,pdfborder={0 0 1},backref=false,colorlinks=true]
 {hyperref}
\hypersetup{
 linkcolor=magenta,urlcolor=blue,citecolor=blue,pdfstartview={FitH},hyperfootnotes=false}

\makeatletter

\usepackage{textcomp}
\usepackage{epstopdf}

\usepackage{amsfonts}

\pdfpageheight\paperheight
\pdfpagewidth\paperwidth

\@ifundefined{textcolor}{}{%
 \definecolor{BLACK}{gray}{0}
 \definecolor{WHITE}{gray}{1}
 \definecolor{RED}{rgb}{1,0,0}
 \definecolor{GREEN}{rgb}{0,1,0}
 \definecolor{BLUE}{rgb}{0,0,1}
 \definecolor{CYAN}{cmyk}{1,0,0,0}
 \definecolor{MAGENTA}{cmyk}{0,1,0,0.2}
 \definecolor{YELLOW}{cmyk}{0,0,1,0}
}

\usepackage{xcolor}\usepackage{soul}
\definecolor{blue}{rgb}{0,0,1}
\definecolor{red}{rgb}{1,0,0}
\definecolor{green}{rgb}{0,1,0}

\usepackage{soul}

\newcommand{\ketbra}[2]{|#1\rangle\!\langle#2|}

\begin{document}
\title{Nonlinearity-Enhanced Continuous Microwave Detection Based on Stochastic Resonance}
\author{Kang-Da Wu}
\affiliation{CAS Key Laboratory of Quantum Information, University of Science and Technology of China, \\ Hefei 230026, People's Republic of China}
\affiliation{CAS Center For Excellence in Quantum Information and Quantum Physics, University of Science and Technology of China, Hefei, 230026, People's Republic of China}
\author{Chongwu Xie}
\affiliation{CAS Key Laboratory of Quantum Information, University of Science and Technology of China, \\ Hefei 230026, People's Republic of China}
\affiliation{CAS Center For Excellence in Quantum Information and Quantum Physics, University of Science and Technology of China, Hefei, 230026, People's Republic of China}
\author{Chuan-Feng Li}
\affiliation{CAS Key Laboratory of Quantum Information, University of Science and Technology of China, \\ Hefei 230026, People's Republic of China}
\affiliation{CAS Center For Excellence in Quantum Information and Quantum Physics, University of Science and Technology of China, Hefei, 230026, People's Republic of China}
\affiliation{Hefei National Laboratory, University of Science and Technology of China, Hefei 230088, People's Republic of China}
\author{Guang-Can Guo}
\affiliation{CAS Key Laboratory of Quantum Information, University of Science and Technology of China, \\ Hefei 230026, People's Republic of China}
\affiliation{CAS Center For Excellence in Quantum Information and Quantum Physics, University of Science and Technology of China, Hefei, 230026, People's Republic of China}
\affiliation{Hefei National Laboratory, University of Science and Technology of China, Hefei 230088, People's Republic of China}
\author{Chang-Ling Zou}\email{clzou321@ustc.edu.cn}
\affiliation{CAS Key Laboratory of Quantum Information, University of Science and Technology of China, \\ Hefei 230026, People's Republic of China}
\affiliation{CAS Center For Excellence in Quantum Information and Quantum Physics, University of Science and Technology of China, Hefei, 230026, People's Republic of China}
\affiliation{Hefei National Laboratory, University of Science and Technology of China, Hefei 230088, People's Republic of China}
\author{Guo-Yong Xiang}\email{gyxiang@ustc.edu.cn}
\affiliation{CAS Key Laboratory of Quantum Information, University of Science and Technology of China, \\ Hefei 230026, People's Republic of China}
\affiliation{CAS Center For Excellence in Quantum Information and Quantum Physics, University of Science and Technology of China, Hefei, 230026, People's Republic of China}
\affiliation{Hefei National Laboratory, University of Science and Technology of China, Hefei 230088, People's Republic of China}

\begin{abstract}  In practical sensing tasks, noise is usually regarded as an obstruction to better performance and will degrade the sensitivity. Fortunately, \textit{stochastic resonance} (SR), a counterintuitive concept, can utilize noise to greatly enhance the detected signal amplitude. Although fundamentally important as a mechanism of weak signal amplification, and has been continually explored in geological, biological, and physical science, both theoretically and experimentally, SR has yet to be demonstrated in realistic sensing tasks. Here we develop a novel SR-based nonlinear sensor using a thermal ensemble of interacting Rydberg atoms. With the assistance of stochastic noise (either inherently in the system or added externally) and strong nonlinearity in the Rydberg ensembles, the signal encoded in a weak MW field is greatly enhanced (over 25 dB). Moreover, we show that the SR-based atomic sensor can achieve a better sensitivity in our system, which is over 6.6 dB compared to a heterodye atomic sensor. Our results show potential advantage of SR-based MW sensors in commercial or defense-related applications.
\end{abstract}


\maketitle

\section{Introduction}


Rydberg atoms provide an ideal platform for linking separated domains of electromagnetic waves: the radio frequency range and optical radiation, due to their large electric dipole moment between Rydberg energy levels~\cite{gallagher2006rydberg}. This distinct property makes Rydberg atom technology a building blocks in a wide variety of applications, such as microwave (MW)-to-optical conversion~\cite{borowka_continuous_2023,hui2022conversion}, terahertz imaging~\cite{wade2018terahertz,chen_terahertz_2022}, and MW electrometry~\cite{jing2020atomic,sedlacek2012microwave,PhysRevLett.111.063001,PhysRevApplied.5.034003,PhysRevA.101.053432,photonics9040250}. Up-to-date, most of the above applications utilize the linear response regime of the Rydberg atoms to external fields. In the case of practical ambient sensing or communication task, the atomic system will inevitably interact with surrounding stochastic noise. In the linear response regime, the noise and useful signal simultaneously enter the system and make the output deteriorate in terms of reduced efficiency or quality. Independent methods have been proposed to suppress ambient noise, such as using a noise filter or screening, at the expense of increasing the system's complexity. 

Recently, there is growing evidence that the strong nonlinearity induced by many body effects in Rydberg atoms may provide a novel method for enhanced sensing of external parameters~\cite{NatureSqueezingYejun,NPDing,Wang:23}. 
However, the pursuit of nonlinearity-induced enhancement in metrology in practical sensing tasks faces a variety of hurdles, as the continuous detection of external fields places stringent demands on the experimental capabilities of controlling and harnessing the non-equilibrium dynamics of the Rydberg system in the presence of systematic or external noise.

Counterintuitively, harnessing the strong nonlinearity may provide a novel method for reversing the negative effect of noise in sensing tasks. More precisely, the atomic ensemble will turn into a nonlinear thresholding device exhibiting bistability, which will collaborate with the external or intrinsic noise, attaining signal-noise synchronization and a greatly enhanced output signal amplitude. This phenomenon, termed \textit{stochastic resonance} (SR)~\cite{RevModPhys.70.223,PhysRevA.39.4854,ThomasWellens2004}, reveals the mechanics of harnessing certain degree of noise in a specific nonlinear system to improve the detection performance instead of hindering it, and has been widely applied in explaining certain phenomenon in different time and spacial scales, such as geological~\cite{doi:10.3402/tellusa.v34i1.10782}, biological~\cite{hanggi2002stochastic,mori2002noise,simonotto1997visual}, and physical science~\cite{PhysRevA.74.013817,SILVA2021105558,FAUVE19835,gingl2001high}. 
The conditions for a system to exhibit SR are well known and fairly robust~\cite{RevModPhys.70.223}. First among these is an energy threshold. Second, there needs to be an applied subthreshold modulation, which causes the system to cross the threshold with the assistance of a source of noise. The above requirements can be achieved using an ensemble of strongly interacting Rydberg atoms, even at room temperature or higher working temperature, instead of carefully engineered systems.

In this work, we develop a novel MW sensor based on strong nonlinearity induced SR in a rubidium vapor cell~\cite{carr2013nonequilibrium,weller2016charge,ding2020phase,de2016intrinsic}. By experimentally using a strong noisy MW field, we can amplify the detected signal amplitude (encoded in a weak MW field) over 25 dB compared with traditional sensor with a linear response, which is also in good agreement with a qualitative model working in the approximation of mean field theory~\cite{carr2013nonequilibrium}. Moreover, the advantages in MW detection of using SR is demonstrated by an enhancement of 6.6 dB in sensitivity compared to heterodyne atomic MW sensor in our system. Our work shows the ability to overcome the negative effect of sthochastic noise with the assistance of many body interactions-induced strong nonlinearity, which paves the way towards developing novel nonlinear and ambient atomic sensors.

\begin{figure*}
	\centering
	\includegraphics[scale=0.21]{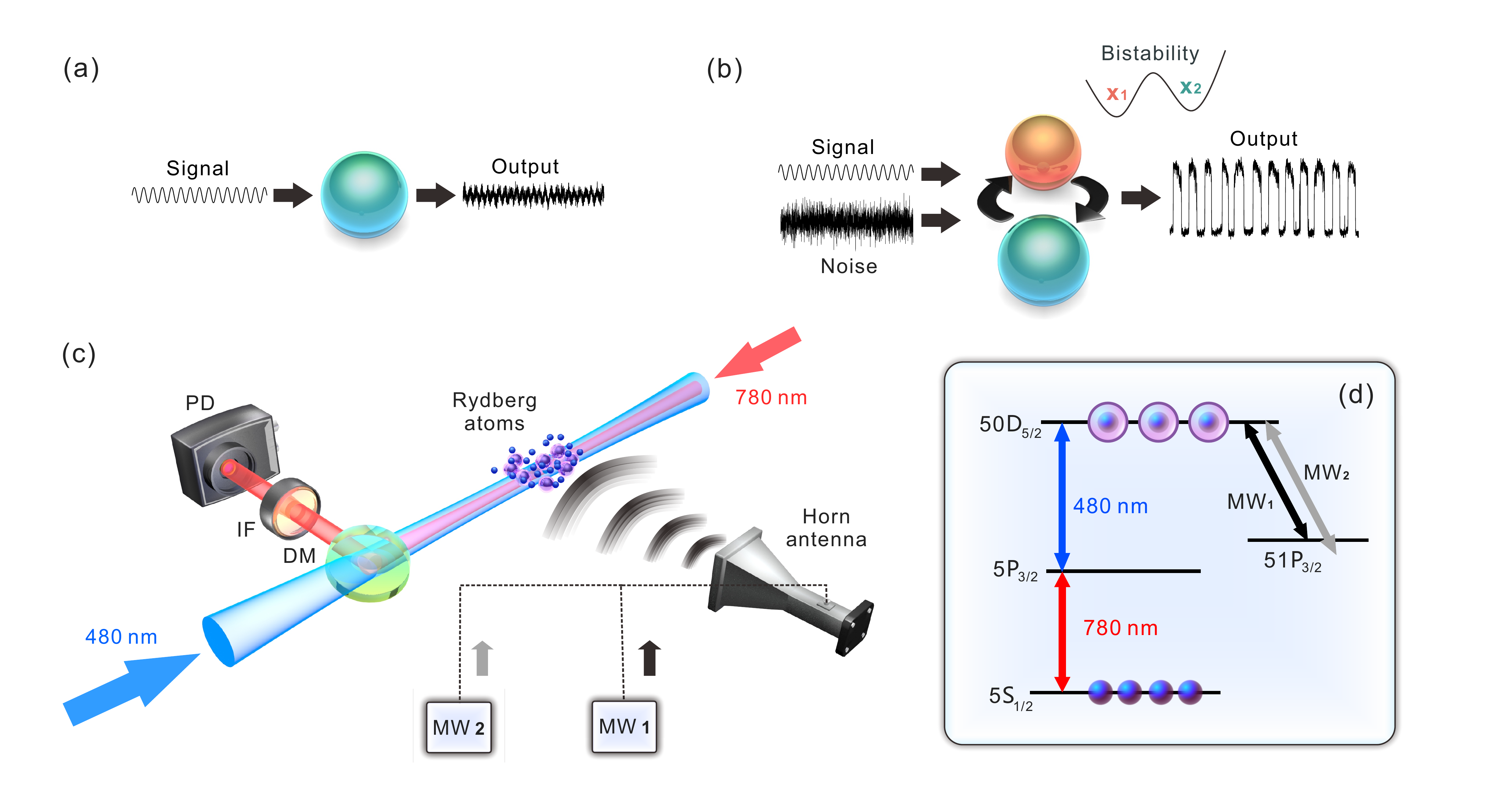}
	\caption{\label{fig:exp}
		\textbf{Conceptual and experimental scheme}. (a,b) Conceptual scheme: (a) the output of a conventional sensor is linearly dependent on the input signal contaminated by inherent noise; (b) a SR sensor can significantly enhance the amplitude of detected signal by externally added white noise. (c,d) Experimental setup and energy level used for the SR experiments. The frequency of probe light (780 nm, horizontally polarized) is stabilized to the transition $5S_{1/2}(F = 2)\leftrightarrow5P_{3/2}(F = 3)$, and overlapped with a counter propagating coupling laser (480.25 nm, horizontally polarized) resonant with the $5P_{3/2}(F = 3)\leftrightarrow50D_{5/2}$ transition inside of the vapor cell contained in a $8\times 8\times 8 $mm$^3$ glass vapor cell heating at a temperature of $69^\circ$C in a ceramic oven. Both lasers are locked to a high finesse cavity using Pound-Drever-Hall (PDH) technique~\cite{10.1119/1.1286663}. The microwave (MW) field contains two frequencies, labeled  "1" and "2", and can be tuned experimentally. The polarization of the MW field is set to the same polarization with both lasers. Meaning of abbreviations: PD, Photoelectric detector; DM, dichroic mirror; MW, microwave.
	}
\end{figure*}

\section{Results}

Figures~\ref{fig:exp}(a,b) schematically illustrate the difference between the conventional and the SR-based MW sensor. In a conventional sensors, the presence of noise  will degrade the output signal-to-noise ratio (SNR). In the case of SR sensor, we consider a nonlinear system that exhibits bistability (threshold), i.e., contains two experimentally accessible stable states ($x_1$, $x_2$), and generates distinguished output for the two states. At specific conditions, critical responses of the system state and a certain amount of perturbation on the system might stimulate the transition between two states. 
The combination of weak signal and certain input noise to the system might effectively activate the transitions between the two states and thus amplify the input signal. For instance, when a weak periodic excitation with a strength below the threshold is subjected to the system, the driving is too weak to stimulate the transition, and the system stays in one of the stable states, the response of the system is just similar to the conventional sensor. In contrast, with certain amount of externally added stochastic noises, the inter-states transition occurs, and can become synchronized with the weak input, resulting in significantly enhanced detected signal.

\textbf{Experimental setup}. We realized the SR-based sensor with Rydberg atom ensemble with the experimental setup shown in Fig.~\ref{fig:exp}(c). The Rydberg atom ensemble provides the essential nonlinear optical medium for the microwave-optical interface for optical detection of electric field measurement, and also the nonlinearity for realizing the critical behavior with respect to the microwave input. The Rydberg atoms are realized with a vapor cell containing hot $^{87}$Rb atoms, with external weak probe light (780 nm) and strong coupling light ($480\,\mathrm{nm}$). The microwave signal near-resonant with Rydberg states $|50D_{5/2}\rangle$ and $|51P_{3/2}\rangle$, and could effectively change the atom's susceptibility to the probe light, as shown by the energy level diagram in Fig.~\ref{fig:exp}(d). Here, we use typically $2.8\,\mu$W probe light for the experiments. Both laser are focused to a $1/e^2$ radius about $100\,\mu$m at the center of the cell. The system dynamics is characterized by the probe light transmission detected by a photoelectric detector (PD), $T\propto U_{\mathrm{PD}}$. The microwave signal is coupled to the system through a horn antenna. 

\begin{figure*}
	\centering
	\includegraphics[scale=0.35]{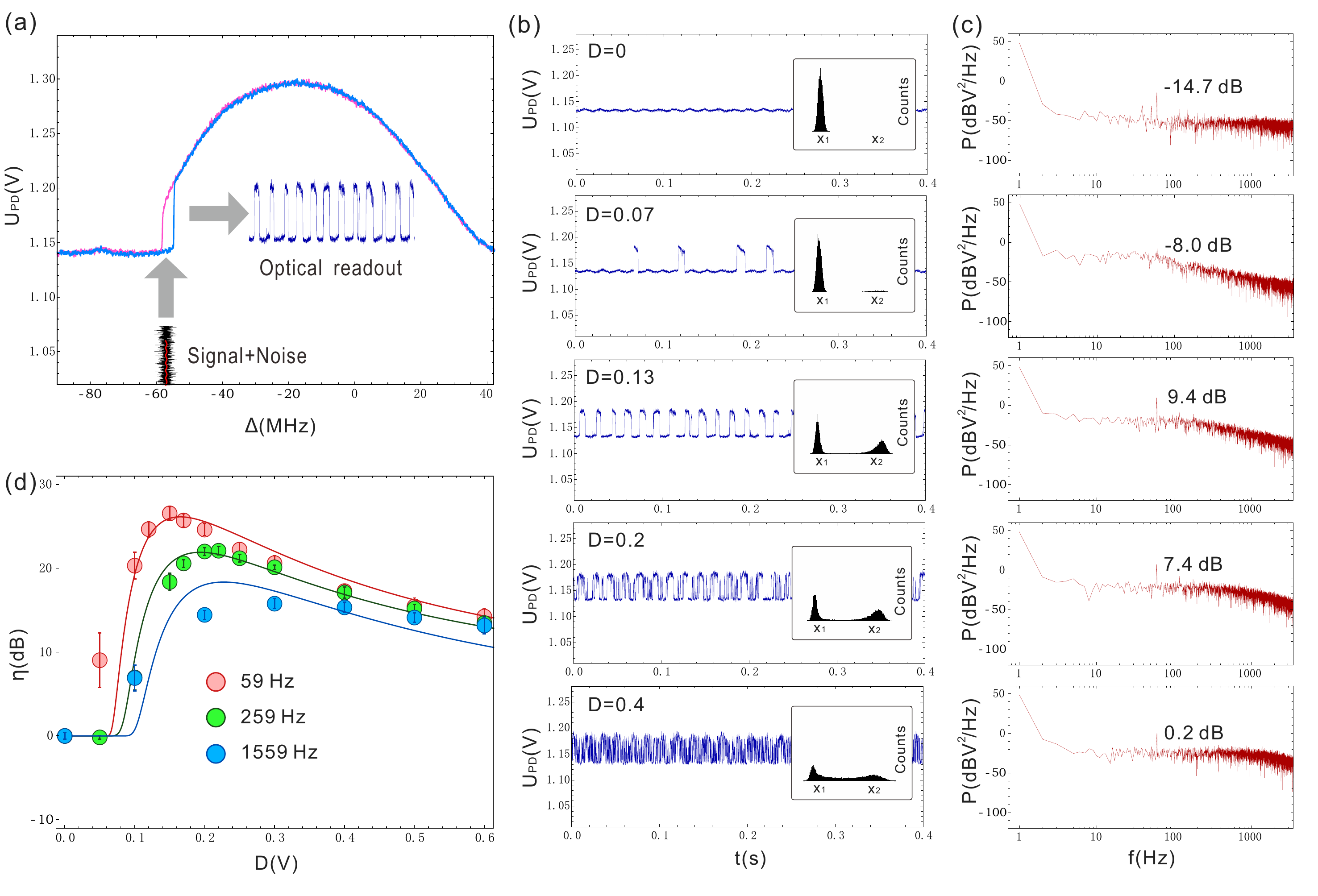}
	\caption{\label{fig:noise-signal}
		\textbf{Experimental results for SR}. In the following experimental results, the power of probe light and coupling light are set to 2.8 $\mu$W and 146 mW, respectively, the detuning of the coupling light is set to around -56 MHz. The noise degree $D$ is the standard variance of applied modulation voltage on MW$_2$,  whose strength is 55.3 mV/cm and frequency is 17.247 GHz. The main results are shown as: (a) MW-induced SR scheme. Both MW$_{1,2}$ will introduce an effective frequency shift $\Delta$ depending on their power. When a modulation is added on the amplitude of MW, the effective loop shift will change according to the modulation voltage. The noise voltage (with controllable strength) is added to the AM of strong MW$_2$ and the sinusoidal signal is added to the AM of weak MW$_1$. Without the noise, the effective $\Delta$ induced by MW$_1$ cannot push the Rydberg population going beyond the hysteresis loop window (threshold). The noise will enhance the amplitude of detected sinusoidal signal with proper noise intensity. (b) time domain optical readout (detected by PD, in units of Volts) and probability distribution around "On"($x_1$) or "Off" ($x_2$) states (inset plot) for different noise degree (ranging from 0 to 0.4). (c) Corresponding Fourier transformation of $U_{\mathrm{PD}}$, shown in units of dB V$^2/$Hz. In (b,c), the modulation signal frequency is 59 Hz. The strength of MW$_1$ is set to 5.5 and 4.4 mV/cm in (b,c) and (d), respectively. (d) The signal gain $\eta$ vs degree of noise $D$ for different modulation frequency. The signal gain is defined as $\eta=10\log_{10} [P_D(f)/P_{D=0}(f)]$where $P_D(f)$ denotes the peak value of the Fourier spectrum at $f$ together with noise degree $D$. Solid lines represent theoretical results (see Supplementary Materials). The error bars in (d) are the standard deviation in the 10 repeated measurements with 1s integral time.
	}
\end{figure*}

\textbf{Experimental demonstration of SR}. Figure~\ref{fig:noise-signal}(a) illustrates the principle of the SR-based electrometry, we apply both signal and noise to the variation in effective coupling laser detunings. The transmission profile $U_{\mathrm{PD}}(\Delta_c)$ is a hysteresis loop, where $\Delta_c$ is the coupling laser detuning. We refer to the higher Rydberg population state as the "On" state (marked as $x_2$) while the lower Rydberg population state as the "Off" state (marked as $x_1$), corresponding to the two states in SR model. Two MWs with frequencies at 17.047 GHz (labeled as "1", resonant) and 17.247 GHz (labeled as "2") are combined and incident at the vapor cell, leading to an additional contribution to $\Delta'_c$ due to AC-Stark shift. 
When both MWs are switched on, we have $\Delta'_{c}= \Delta_1+\Delta_2+\Delta_c$, where $\Delta_{1,2}$ represents MW$_{1,2}$-induced frequency shift. Here, the signal and noise are provided by the variations in $\Delta_1$ and $\Delta_2$, respectively, which are introduced by the amplitude modulation (AM) of MW$_{1,2}$ . Since the power of MW$_1$ is weak, the coherent modulation of MW$_1$ can not induce a effective shift greater than the loop window, thus the periodic readout signal will be the local oscillation in one of two stable states; when introducing stochastic noise in the AM of MW$_2$, the situation is completely different, the large fluctuations in effective shift in $\Delta'_c$ will exceed the threshold, and greatly enhance the detected signal. More experimental details are shown in the Supplementary Materials.

\begin{figure*}
	\centering
	\includegraphics[scale=0.36]{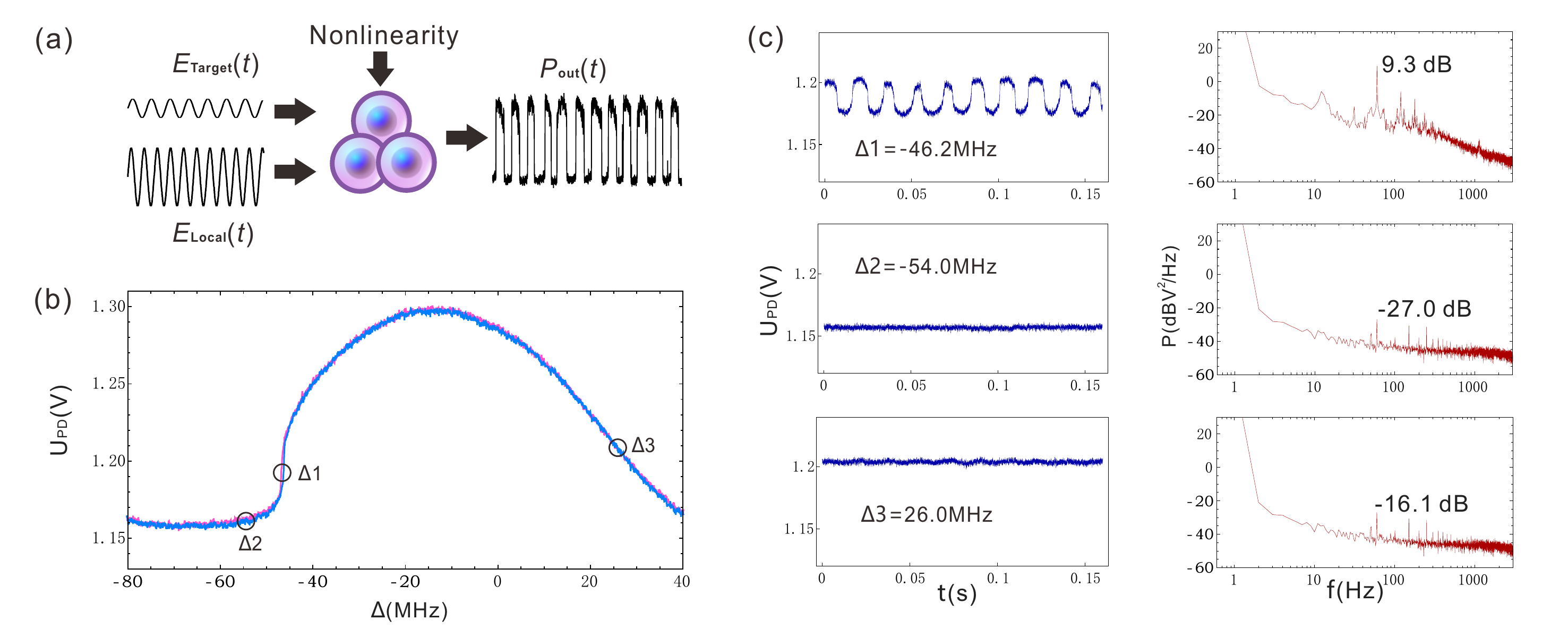}
	\caption{\label{fig:heterodyne}
		\textbf{Experimental results for SR in MW sensing}. In this part, the same experimental setup is used and a comparison between SR sensor and heterodyne sensor is made. The previous MW$_{1(2)}$ is denoted by target (local) MW according with a heterodyne configuration where the beat ($\delta_f=59$Hz) signal between the two fields is detected. Only systematic noise is used instead of externally added modulation. (a) Sensing applications adopting the heterodyning configuration. (b) For a better  performance, we decrease the coupling power to $P_c$=127 mW. The hysteresis loop window is nearly close (with a width of around 1 MHz), the local MW power is  $E_{\mathrm{Local}}$=11.0 mV/cm while scanning $\Delta_c$. (c) Times series of PD readout and corresponding power density spectra for three different coupling detunings $\Delta_c$=-46.2, -54.0, and 26.0 MHz, the target field strength is $E_{\mathrm{Target}}$=391.4$\mu$V/cm. 
	}
\end{figure*}

\begin{figure}
	\centering
	\includegraphics[scale=0.36]{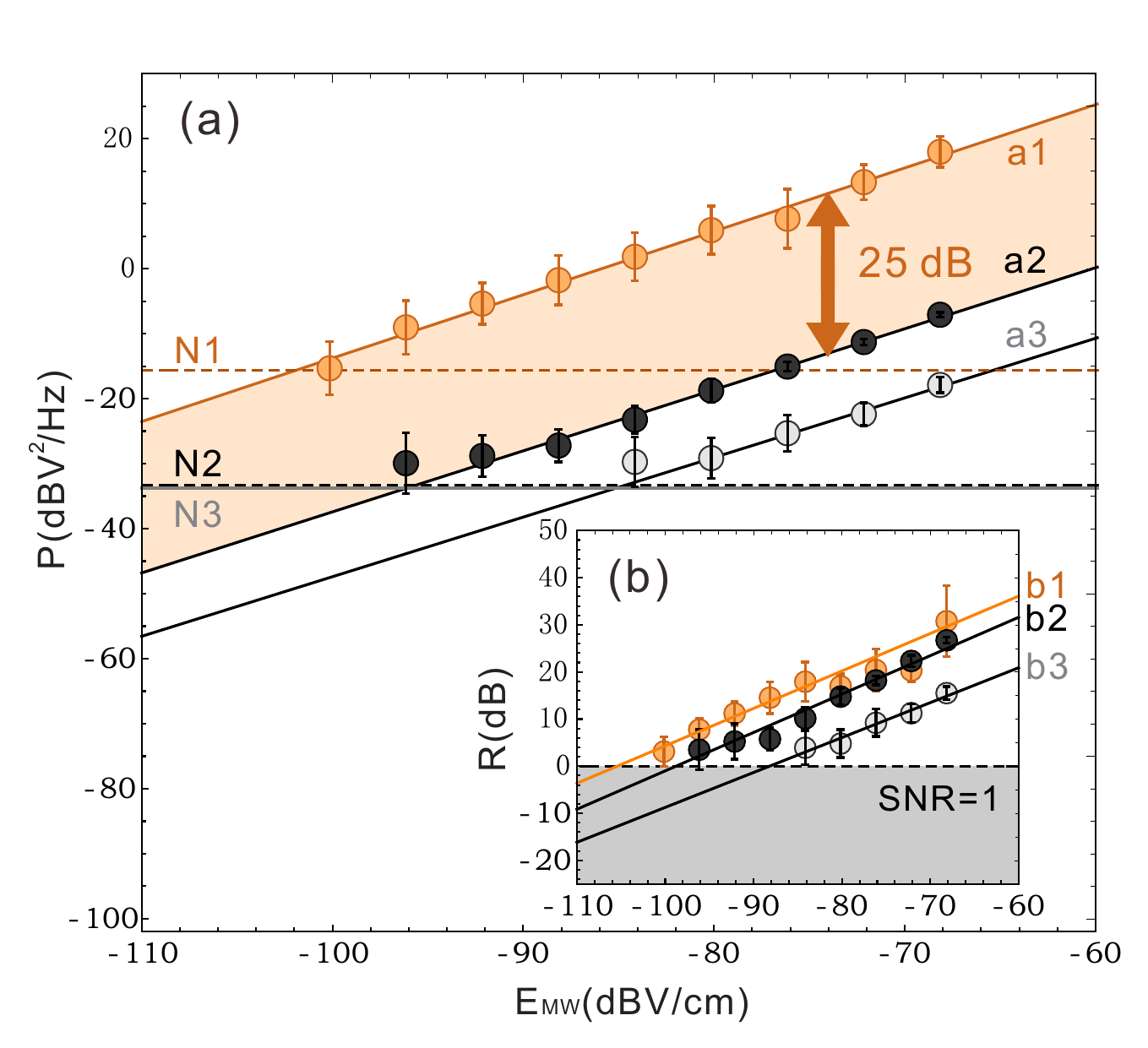}
	\caption{\label{fig:sensor}
		\textbf{SR sensing performance}. Experimental results for MW detection using SR (orange disks) vs non SR case (black and white disks) shown with detected amplitude (a) and SNR (b). In (a), N1-N3 represent the noise strength obtained from the Fourier transformed data. In (b) the SNR is calculated by $R=10\mathrm{log}_{10}[P_D(f)/\bar{P}_D]$ where $P_D(f)$ is the peak value, $\bar{P}_D$ is the estimated average power of noise around $f$ (noise at specific frequencies are not counted, such as 50 Hz), and the error bars in both (a) and (b) are the standard deviation in the 10 repeated measurements with 1s integrating time. The values of horizontal axis reads 20log$_{10}E_{MW}$ (dB V/cm).
	}
\end{figure}

Experimentally, the steps to observing SR are as follows: (i) switch on both MW fields and scan the coupling laser frequency to observe hysteresis loop while locking the frequency of the probe laser; (ii) lock the frequency of the coupling light inside the window with the assistance of the reference cavity; (iii) apply both a periodic signal $s(t)=\sin\left( 2\pi\times f t\right)$ to the AM of MW$_1$, and Guassian white noise to the AM of the MW$_2$ ($E_{MW_2}$=55.3 mV/cm). Here, the noise strength is quantified by the standard variance $D$ (in the units of V, and 1 V corresponds to an AM depth of 100\%) of the applied modulation voltage. Figures~\ref{fig:noise-signal}(b,c) illustrate the experimental results for SR with signal frequency $f=$59Hz with different noise level.
The voltage output of the detector (PD) as a time series is stored digitally, and a fast Fourier transform is performed on the data to obtain the power spectrum. For a better demonstration of SR, we set MW$_1$ to 5.5 mV/cm, which is slightly larger than later experiments. Figure~\ref{fig:noise-signal} (b) shows the time series and residence distribution for different $D$. Each plot is the result of tracked PD output for 0.4s with a sampling rate of 50 kHz. When $D$ is below the threshold (around 0.06 for $f$=59 Hz), the residence distribution admits an approximately Gaussian shape centering at one of the two states. As $D$ is increased beyond the threshold, signal-noise synchronization takes place, and the residence distribution admits approximately two Gaussian shape centering at both the "On" and "Off" states. In the meantime, the line shape of two Gaussian distribution broadens due to the enhanced inner-state motions. These results clearly show the basic properties of SR that with a properly amount of white noise, the sub-threshold period signal is detected with the assistance of noise, the lose of signal-noise synchronization happens at an exorbitantly high noise level. 

The SR-based MW sensor is further characterized under different frequencies and noise conditions. We set the strength of MW$_1$ to 4.4 mV/cm, and track the PD output for various signal frequencies when increasing $D$ from 0 to 0.6 V. The experimental results are shown in Fig.~\ref{fig:noise-signal} (d). In particular, the amplification $\eta=10\mathrm{log}_{10}[P_D(f)/P_0(f)]$ at signal frequency $f$ is plotted against $D$, together with a theoretical analysis (solid lines) derived by regarding the dynamics of the Rydberg population as a Brownian particle inside a double well potential (see Supplementary Materials). 
\\

\textbf{Applications in MW sensing.} We now show how our nonlinear atomic sensor has an advantage in weak signal detection in the presence of noise. We adopt a recently widely adopted MW atomic sensor based on heterodyne technology~ \cite{simons2019rydberg,jing2020atomic}, and compare the sensitivity gain of our new sensor with that of a traditional heterodyne sensor. The principle of a heterodyne atomic sensor is illustrated in Fig.~\ref{fig:heterodyne} (a), which contains a target MW field whose strength is to be detected and a strong local MW field. The total field strength felt by the Rydberg atoms can be written as $E_\mathrm{Total}=E_\mathrm{Local}+E_\mathrm{Target}\cos{(\delta_f t + \phi)}$ where $\delta_f$ and $\phi$ are the frequency and phase difference between the local and target MWs, resulting in the experimentally detectable beat note of two fields. Here, we refer the previous MW$_{1,2}$ to as "target" ("local") in accordance with the heterodyne method. 

In ambient sensing tasks based on SR, we need to deal with stochastic noise in the surroundings, and turn the negative effects of noise into positive ones by utilizing nonlinearity. In our laboratory, the effective noise strength (including fluctuations of residual ground magnetic or electric fields) is relatively small, thus we need to decrease the degree of nonlinearity of Rydberg atoms to match the noise strength. Figure~\ref{fig:heterodyne} (b) shows the experimental hysteresis loop with local MW switched on in the case of lower coupling power $P_c=127$ mW, where we can identify the width of the loop window to be around 1 MHz, which is close to the critical point in Rydberg system~\cite{Universal2014}. In this regime, stochastic noise in laboratory environment start to dominate and lead to spontaneous transitions with $\Delta_c$ set to a proper value inside loop window~\cite{ding2020phase,he2020stochastic}.

For a fair comparison, we use the same experimental condition for the implementation of different sensing methods. The local MW, with a strength of 11.0 mV/cm, is set $\delta_f=59$Hz blue detuned to the target one(17.047GHz). We set $\Delta_c=-46.2$MHz (inside the loop window) for SR-based sensing, and $\Delta_c=-54.0$ and $26.0$ MHz (larger slope) for comparison experiments with the heterodyne method. In Fig.~\ref{fig:heterodyne} (c), typical time domain signals and corresponding power spectral density are shown for above three cases. It is clear that when working in the SR regime, stochastic transitions between the two stable states happens and can assists us in enhancing the detection of a weak target field. Figs.~\ref{fig:sensor}(a,b) show the enhancement in detected signal amplitude ($P$) and SNR ($R$) for different target powers, respectively. The detected optical signal amplitude in SR regime (a1) is amplified over 25 dB over the two heterodyne cases (a2,a3). And the SNR is increased by approximately 6.6 dB over the other two two case (b2,b3) using heterodyne method. In the case that the detectable variation in optical signal is comparable to the PD readout noise, SR will offer an enhancement in output SNR since the optical signal is greatly amplified.

\section{Discussion}
An SR-based nonlinear atomic MW sensor is experimentally realized in a thermal ensemble of interacting Rydberg atoms, which can continuously detect a MW field rather than using the nonequilibrium dynamics around the critical point. Experimentally, a significantly enhanced (over 25 dB at a modulation frequency of 59 Hz) amplitude of optically detected signal is observed, and some key features of SR have been demonstrated. Moreover, an improved sensitivity (over 6.6 dB with respect to heterodye sensor) is demonstrated with based on SR. The experimental results have shown that the noise does not need to be inherently in the system but can also be tuned and added externally. This can mark a paradigm shift in future noisy signal processing and sensing design, and is ready to be practically extended to devices used in commercial or defense-related applications. 

The demonstration of SR in Rydberg atomic systems is exciting not only for its inherent physical interest but also because it brings up the possibility of using it to enhance signal processing and electric field sensing. In most atomic sensing settings, the number of atoms and interaction volume play essential roles. In our current SR atomic sensor, however, the home-built coupling laser power is limited to 200 mW, leading to a smaller beam waist shorter cell in order to increase the Rabi frquency of coupling lasers and match the Rayleigh length. The interaction volume are much smaller due to the above reasons (around 450 times smaller than that in Ref.~\cite{jing2020atomic}). We expect a sensitivity better than 100 nV/cm/$\sqrt{\mathrm{Hz}}$ with increased interaction volume and atom numbers. The sensitivity of SR sensor based on hot atoms is also limited by Doppler broadening, atom-atom collision and transient noise. Quantum SR~\cite{PhysRevLett.72.1947} utilizing the atom shot noise, black body radiation, and the vacuum fluctuations of MW fields, provides a potentially achievable sensitivity at the level of several nV/cm/$\sqrt{\mathrm{Hz}}$~\cite{santamariabotello2022comparison}, when other technical noise can be sufficiently suppressed.
\\

{\bf Acknowledgment.} The work at the University of Science and Technology of China was supported by the National Key R\&D Program (Grant No. 2021YFA1402004), the National Natural Science Foundation of China (Grants Nos. 12134014, 11974335, 11574291, and 11774334), the Key Research Program of Frontier Sciences, CAS (Grant No. QYZDYSSW-SLH003), USTC Research Funds of the Double First-Class Initiative (Grant No. YD2030002007) and the Fundamental Research Funds for the Central Universities (Grant No. WK2470000035).\\

K.-D.~W.~and C.~X.~contributed equally to this work. 

\clearpage

\pagebreak
\onecolumngrid

\begin{center}
	\textbf{\large Supplementary Material for Nonlinearity-Enhanced Continuous Microwave Detection Based on Stochastic Resonance}
\end{center}
\setcounter{equation}{0}
\setcounter{figure}{0}
\setcounter{table}{0}
\setcounter{page}{1}
\renewcommand{\theequation}{S\arabic{equation}}
\renewcommand{\thefigure}{S\arabic{figure}}
\renewcommand{\thepage}{S\arabic{page}}
\renewcommand{\thesection}{S\arabic{section}}
\renewcommand{\bibnumfmt}[1]{[S#1]}
\renewcommand{\citenumfont}[1]{S#1}

\section*{S1. Models}

\subsection{Effective potential}
We now provide another understanding of SR in MW-Rydberg system by considering effective potential $U$ for the Rydberg system with order parameter as the Rydberg population $\rho_{rr}$. In particular, we treat each atom in the vapor as a two-level atom, and there are interaction between light and atoms as well as many body interactions between atoms, electrons, or ions. Then the Hilbert space of the whole atomic ensemble  will be $2^n$-dimensional (n is the number of atoms), but it becomes quite difficult to directly solve in such a large space. So we use the mean field theory (MFT) to simplify the formalism of many body interactions. The Hamiltonian of the light-atom interaction model with the MFT under rotating wave approximation (RWA) is given by \cite{PhysRevLett.101.250601,carr2013nonequilibrium}:

\begin{align}\label{eq:hamitonian}
	H=-\Delta \hbar \ketbra{r}{r} +\dfrac{\Omega\hbar}{2}\left( \ketbra{g}{r}+\ketbra{r}{g}\right) -V_{int}\rho_{rr}\hbar\ketbra{r}{r}, 
\end{align}
where $ | g \rangle$ is the ground state and $| r \rangle$ is the Rydberg state, $\Delta$ is the detuning of light, $\Omega$ is the Rabi frequency, $\rho$ is the density matrix, $V_{int}\rho_{rr}$ is the effective detuning caused by the Rydberg interaction and the spontaneous ionization  which dependent on the atomic number density and the principle number of the Rydberg level. In the Eq\eqref{eq:hamitonian}, the 1st term is the atom itself energy, the 2nd term is the atom-light interaction term, the 3rd term is the Rydberg interaction term which comes from the MFT. The master equation is given by:
\begin{align}\label{eq:masterquation}
	\dot{\rho}=\dfrac{1}{i\hbar}[H,\rho]+\mathcal{L}(\rho),
\end{align}
where $\mathcal{L}(\rho)$ account for the incoherent decay caused by environments and the system. We can write the eq\eqref{eq:masterquation} in number not the operator:
\begin{align}\label{eq:masterex1}
	\dot{\rho}_{gr}&=-i\left(\Delta\rho_{gr}-\dfrac{\rho_{gg}\Omega}{2}+\dfrac{\rho_{rr}\Omega}{2}\right)-\dfrac{(\Gamma+\gamma)\rho_{gr}}{2},\\
	\dot{\rho}_{rr}&=-i\left(\dfrac{\rho_{gr}\Omega}{2}-\dfrac{\rho_{rg}\Omega}{2}\right)-\Gamma\rho_{rr},\\
	\rho_{gg}&=1-\rho_{rr},\\
	\rho_{rg}&=\rho_{gr}^{*}.
\end{align}
Here $\Gamma$ is the decay rate from $|r\rangle$ to $|g\rangle$ which is mainly contributed by the spontaneous emission, $\gamma$ is the decay of coherent term which is caused by the other decoherence mechanics.
Because of long lifetime of Rydberg atom and the enhanced dephasing caused by Rydberg interaction~\cite{PhysRevLett.110.123001} and Doppler boarding, we make assumption that in  the coherent terms ($\rho_{rg},\rho_{gr}$) evolve much faster than the population terms($\rho_{rr},\rho_{gg}$)which means the coherent terms quickly reaches equilibrium while the population terms is still evolving i.e. $\gamma \gg \Gamma$:
\begin{align}\label{eq:masterex-1}
	\dot{\rho}_{rg}=\dot{\rho}_{gr}=0.
\end{align}
Combine Eqs.\eqref{eq:masterex1}-\eqref{eq:masterex-1},we can get:
\begin{align}\label{eq:systemev}
	\dot{\rho}_{rr}= -\rho _{rr}\Gamma+\dfrac{(\Gamma+\gamma)\left( 1-2\rho_{rr}\right) \Omega ^{2}}{(\Gamma+\gamma)^2+4\left( \Delta +V_{\mathrm{int}}\rho _{rr}\right) ^{2}}.
\end{align}
In the words of synergetics~\cite{Haken_1977}, we refer $\rho_{rr}$ to the order parameter of the system and other parameters are salved by $\rho_{rr}$. We can define an effective potential $U_{\mathrm{eff}}$ for our system:
\begin{align}
	\dot{\rho}_{rr}=-\frac{\partial\, U_{\mathrm{eff}}(\rho_{rr};\Omega,\Delta)}{\partial\, \rho_{rr}},
\end{align}
where $U_{\mathrm{eff}}(\rho_{rr};\Omega,\Delta)$ is the effective potential with order parameter $\rho_{rr}$.
If we replace $\rho_{rr}$ with the generalized position of a Brownian particle $x\equiv\rho_{rr}$ and $\Gamma_c=\Gamma+\gamma$, we have
\begin{equation}\label{eq:effectivepotential}
	\begin{aligned}
		U_{\mathrm{eff}}(x)&=-\frac{\Omega^2(V_{\mathrm{int}}+2\Delta_0)}{2V^2_{\mathrm{int}}}\arctan\frac{2\Delta(x)}{\Gamma_c}\\&+\frac{x^2\Gamma}{2}+\frac{\Omega^2\Gamma_c}{4V^2_{\mathrm{int}}}\mathrm{ln}\left[\Gamma_c^2+4\Delta^2(x)\right].
	\end{aligned}
\end{equation}
The above equation is the effective potential for a Brownian particle whose position $x$ is  determined the Rydberg population $\rho_{rr}$. In our hot Rydberg system, we expect this effective potential can help us understand the physics of SR in a Rydberg atomic system qualitatively.

\subsection{General framework for SR in Rydberg system}
From above section, we simplify the state of whole system to one single parameter($\rho_{rr}$) and we know it's evolution form(eq\eqref{eq:systemev}), thus we can define an effective potential $U_{eff}$ for our system:
\begin{align}
	\dot{\rho}_{rr}=-\frac{\partial\, U_{\mathrm{eff}}(\rho_{rr};\Omega,\Delta)}{\partial\, \rho_{rr}},
\end{align}


Eq.~(\ref{eq:effectivepotential}) implies that $U_{\mathrm{eff}}$ admit a asymmetric double well potential, and the corresponding two local minimums are the two experimentally accessible stable states in SR. The presence of noise (through one of the parameters) injected to the system will cause inter-well motions between the two stable states. Note that the potential is asymmetric which means the noise-induced transitions $W_{1\rightarrow2}\neq W_{2\rightarrow1}$ for $s(t)=0$. In our experiments, the noise is added to effective $\Delta_c$ we consider only the first order of $\xi(t)$, i.e., we have \begin{equation}\Delta\approx\Delta_c+\alpha_r E_0^2\left[1+2\xi(t)\right]=\Delta_0+\xi(t),\end{equation}
where $\Delta_0=\Delta_c+\alpha_rE_0^2$, and $\zeta(t)=2\alpha_rE_0^2\xi(t)$.

Now considering effective detuning induced by the periodic modulation $s(t)$, we are ready to expand Eq.~(\ref{eq:systemev}) as

\begin{equation}\label{eq:SRequation}
	\begin{aligned}
		&\dot{x}=-\frac{\partial U_\mathrm{eff}(x,\,\Delta_0)}{\partial x}+K\,s(t)-\frac{8(2x-1)(V_\mathrm{int}x+\Delta_0)\Gamma_c}{\left[\Gamma_c^2+4\Delta^2(x)\right]^2}\xi(t).
	\end{aligned}
\end{equation}
Here $s(t)=A_0\cos \omega_s t$, $K=K(x,\Delta_0,\Omega)\propto \Omega_{MW_1}$ is a coefficient, and the variance of the magnitude of $\zeta(t)$ is no larger than $4$MHz. Since the strength of noise is also considered weak compared to many other parameters such as $\Delta_0$ and $V_\mathrm{int}x$and the coefficients of second and third order of $\zeta$ are smaller than the first order, we just drop them. We can further simplify the noise parameter as the mean value around the local minimum. Since the overdamped motion of the Brownian particle in $U_{\mathrm{eff}}(x)$ in the presence of noise and periodic forcing can be written as
\begin{equation}
	\dot{x}=-\frac{\partial U_{\mathrm{eff}}(x)}{\partial x}+K\,s(t)+D\xi(t),
\end{equation}
 where $D=D(x,\Delta_0)$ can be obtained directly from Eq.~(\ref{eq:SRequation}). Here $A_0$ corresponds to both the modulation depth and MW$_1$ strength.

We can further simplify the model to a two-state dynamics where the system is described by a discrete random dynamical variable $x$ that adopts two possible values $x_1$ and $x_2$. We  assume that the probabilities for two states are $n_1(t)$ and $n_2(t)$, satisfying $n_1(t)+n_2(t)=1$. In the presence of noise with strength $D$, the master equation governing the evolution of $n_1(t)$ (and similarly for $n_2(t) =1-n_1(t)$), induced by purely the noise, reads
\begin{equation}\label{eq:masterequation}
	\dot{n}_1(t)=-\dot{n}_2(t)=W_2(t)n_2(t)-W_1(t)n_1(t),
\end{equation}
where $W_1$ and $W_2$ denote the $D$-dependent Kramers escaping rate which can be expressed as 
\begin{equation}\begin{aligned}
		&W_1(t)=\frac{\sqrt{|U''(x_1)U''(x_b)|}}{2\pi}\exp\left[-\frac{U(x_b)-U(x_1)}{D^2}\right]\\
		&W_2(t)=\frac{\sqrt{|U''(x_2)U''(x_b)|}}{2\pi}\exp\left[-\frac{U(x_b)-U(x_2)}{D^2}\right],
	\end{aligned}
\end{equation}
where $x_{1,2}$ denotes the locations (Rydberg population) of two minimum potential and $x_b$ denotes the barrier location. We note that $x_i$ satisfy $\partial U/\partial x=0$.

Following Eq.~(\ref{eq:masterequation}), we have
\begin{align}
	&n_{1,2}(t)=g(t)\left[n_{1,2}(t_0)+\int_{t_0}^{t}W_{2,1}(\tau)g^{-1}(\tau)d\tau\right]\\
	&g(t)=\exp\left(-\int_{t_0}^{t}\left[W_1(\tau)+W_2(\tau)d\tau\right]\right).
\end{align}

Applying a weak periodic signal $s(t)=A_0\cos\omega_st$ as in Eq.~(\ref{eq:SRequation}), up to first order on its amplitude (assumed to be small) the transition rates are periodically modulated, and can be expanded as
\begin{equation}\begin{aligned}
		&W_1(t)\approx r_1+k_1 A_0\cos\omega_st\\
		&W_2(t)\approx r_2+k_2 A_0\cos\omega_st,
	\end{aligned}
\end{equation} 
where $r_{1,2}$ denotes the transition rate without the signal $s(t)$, and $k_{1,2}=\frac{\partial W_{1,2}(t)}{\partial s(t)}|_{s(t)=0}$. By denoting $r=r_1+r_2$ and $k=k_1+k_2$, We have $W_1(t)+W_2(t)=(r_1+r_2)+(k_1+k_2)A_0\cos\omega_st=r+kA_0\cos\omega_st$, we can further express $g(t)$ as

\begin{equation}
	g(t)=\exp\left(-\left[r(t-t_0)+A_0k\frac{\sin\omega_st-\sin\omega_st_0}{\omega_s}\right]\right)\approx\exp[-r(t-t_0)]\frac{\omega_s+kA_0(\sin\omega_st-\sin\omega_st_0)}{\omega_s}.
\end{equation}

Now we are ready to integrate Eq.~(\ref{eq:masterequation}) to the first order of $A_0\cos\omega_st$ as

\begin{equation}
\begin{aligned}
		n_{1}(t;x_0,t_0)&=\frac{r_2}{r}+A_0\frac{(k_2 r_1-k_1 r_2)(r\cos\omega_st+\omega_s\sin\omega_st)}{r(r^2+\omega_s^2)}\\
		&+e^{-r(t-t_0)}\left[A_0\frac{(-k_2 r_1 + k_1 r_2)\cos\omega_st}{r^2+\omega_s^2}+\frac{(-r_2+rn(x_0,t_0))(\omega_s-A_0k\sin\omega_st)}{r\omega_s}\right]\\
		&+A_0e^{-r(t-t_0)}\left[\frac{[kr(-r_2+rn(x_0,t_0))+(k_2[-1+n(x_0,t_0)]+k_1n(x_0,t_0))]\sin\omega_st}{\omega_s(r^2+\omega_s^2)}\right].
\end{aligned}
\end{equation}

We now proceed to calculate the time-dependent response $\langle x(t)|x_0,t_0\rangle=\int x P(x,t|x_0,t_0)dx$ where $P(x,t|x_0,t_0)=n_1(t)\delta(x-x_1)+n_2(t)\delta(x-x_2)$.
Thus we obtain
\begin{equation}
	x(t)=x_1n_1(t)+x_2n_2(t)=x_2+(x_1-x_2)n_1(t),
\end{equation}
Asymptotically, the
memory of the initial conditions gets lost and
\begin{equation}\begin{aligned}
		&\lim_{t_0\rightarrow-\infty}\langle x(t)|x_0,t_0\rangle=x_2+(x_1-x_2)\lim_{t_0\rightarrow-\infty}n_{1}(t;x_0,t_0)\\
		&=x_2+(x_1-x_2)\left[\frac{r_2}{r}+\frac{A_0(k_2r_1-k_1r_2)\cos(\omega_st-\phi)}{r\sqrt{r^2+\omega_s^2}}\right]\\
		&=\frac{x_1r_2+x_2r_1}{r}+\frac{(x_1-x_2)(k_2r_1-k_1r_2)\cos(\omega_st-\phi)}{r\sqrt{r^2+\omega_s^2}}A_0,
\end{aligned}\end{equation}
where $\phi=\arctan(\omega_s/r)$. Since in experiments we perform the fast Fourier transformation, the peak value at $\omega_s$ can be estimated as the square of the module of amplitude of the periodic system response:
\begin{equation}
	P(\omega_s)=2\pi\frac{A_0^2(x_1-x_2)^2(r_2k_1-r_1k_2)^2}{2r^2(r^2+\omega_s^2)}
\end{equation}

We now move to calculate the power spectrum. In particular, we need to estimate the correlation function $\langle x(t+\tau)x(t)|x_0,t_0\rangle$ as

\begin{equation}\begin{aligned}
		\langle x(t+\tau)x(t)|x_0,t_0\rangle&=x_1^2n_1(t+\tau|x_1,t)n_1(t|x_0,t_0)+x_1x_2n_1(t+\tau|x_2,t)n_2(t|x_0,t_0)\\&+x_1x_2n_2(t+\tau|x_1,t)n_2(t|x_0,t_0)+x_2^2n_2(t+\tau|x_2,t)n_2(t|x_0,t_0).
	\end{aligned}
\end{equation}
Since we consider the time-averaged $\langle x(t+\tau)x(t)|x_0,t_0\rangle$ in time scale of the period $T=2\pi/\omega_s$, which can be estimated as
\begin{equation}
	\begin{aligned}
		&\langle\langle x(t+\tau)x(t)|x_0,t_0\rangle\rangle_t=\frac{\omega_s}{2\pi}\int_{0}^{\frac{2\pi}{\omega_s}}\langle x(t+\tau)x(t)|x_0,t_0\rangle dt\\
		&=\frac{(r_2x_1+r_1x_2)^2}{r^2}+e^{-r|\tau|}\frac{r_1r_2(x_1-x_2)^2}{\omega_sr^2}+A_0^2\frac{(k_2r_1-k_1r_2)^2(x_1-x_2)^2\cos\omega_s\tau}{2r^2(r^2+\omega_s^2)}\\
		&+A_0^2e^{-r|\tau|}\frac{(k_2r_1-k_1r_2)(x_1-x_2)\left[r(k_1 x_1 + k_2 x_2)\omega_s-k(r_1 x_1 + r_2 x_2)(\omega_s \cos\omega_s\tau  + r \sin\omega_s\tau)\right]}{2r^2\omega_s^2(r^2+\omega_s^2)},
	\end{aligned}
\end{equation}
and denoting $R_0=\frac{(r_2x_1+r_1x_2)^2}{r^2}$, we can further get the Fourier Transformation of $\langle\langle x(t+\tau)x(t)|x_0,t_0\rangle\rangle_t-R_0$ as 
\begin{equation}
	S(\omega)=\int\langle\langle  (x(t+\tau)x(t)|x_0,t_0\rangle\rangle_t-R_0)e^{-i\omega\tau}d\tau.
\end{equation}
We finally obtain:
\begin{equation}
	S(\omega)=\sqrt{\frac{2}{\pi}}\left[\frac{(x_1-x_2)^2r_1r_2}{r(r^2+\omega^2)}+O(A_0^2)\right]+\sqrt{\frac{\pi}{2}}\frac{ A_0^2(x_1-x_2)^2(r_2k_1-r_1k_2)^2}{2r^2(r^2+\omega^2)}\left[\delta(\omega-\omega_s)+\delta(\omega+\omega_s)\right],
\end{equation}
and the SNR can be calculated as:
\begin{equation}
	R=\frac{A_0^2\pi(r_2k_1-r_1k_2)^2}{4r_1r_2r}.
\end{equation}
Note that $P(\omega_s)=0$ when $D$=0 theoretically. However, in experiments, the dynamics cannot be completely simplified to a two-state model, as there is an nonzero osillation at frequency $\omega_s$ when added noise $D=0$. Taking this into account, we modify the average response to 
\begin{equation}\begin{aligned}
	x(t)=\frac{x_1r_2+x_2r_1}{r}+x_0\cos(\omega_st-\phi_0)+\frac{(x_1-x_2)(k_2r_1-k_1r_2)\cos(\omega_st-\phi)}{r\sqrt{r^2+\omega_s^2}}A_0,
\end{aligned}\end{equation}
where $x_0$ corresponds to the nonzero amplitude when considering the inner state motion. Thus the amplification factor can be modelled as $\eta=\mathrm{log}_{10}[1+k_p A_0^2(x_1-x_2)^2(r_2k_1-r_1k_2)^2/2r^2/(r^2+\omega^2)]$, where $k_p$ is related to $x_0$ and $\omega_s$ and need to be fit in the experiments. Accordingly, the SNR $R$ also needs to be modelled considering inner states motions, i.e., $R=A_0^2\pi(r_2k_1-r_1k_2)^2/(4r_1r_2r)+k_rr(r^2+\omega^2)/(r_1r_2)$ with experimentally fitting $k_r$.

For fitting the experimental data, we set $\Delta_0=-52$ MHz, $\Omega=5.0$MHz, $\Gamma=450$kHz, $\gamma=5.0$MHz, and $V=126.0$ MHz. Here, the proportional coeficient $k_pA_0$ varies with varying target frequency $\omega_s$ as our system has limited bandwidth, leading to different responsibility at different frequency. More precisely, $k_pA_0$ equals 0.14, 0.082, 0.05 for $\omega_s$=59 Hz, 259 Hz, and 1559 Hz, respectively. In all experimental fittings, we consider the average noise between $x_{1,2}$ and $x_0$. 


\section*{S2. Experimental details}
\textbf{Generation of the probe and coupling light.}
A continues 780 nm light (Toptica DLC pro) whose frequency is stabilized to the transition of $5S_{1/2}(F = 2)\leftrightarrow5P_{3/2}(F = 3)$ with the assistance of a high finesse cavity ($\mathcal{F}\approx$15000), is focused to a $1/e^2$ radius $100\mu$m at the centre of the cell and is overlapped with a counter propagating coupling laser (480.25 nm, $1/e^2$ radius $80\mu$m) resonant with the $5P_{3/2}(F = 3)\leftrightarrow50D_{5/2}$ transition inside of the vapor cell. The 480 nm coupling light is derived from a homemade frequency doubling system using second harmonics generation (SHG) from a 1.4 W continues laser (Toptica TA pro) with a central wavelength of 960 nm. To increase the efficiency of SHG and the power of blue light, a bow-tie cavity consists of four mirrors is used. In particular, the 960 nm laser (Toptica TA pro) is mode matched and enters the bow-tie cavity through a partial reflective plane mirror with a transmissivity of $10\%$, one of the curved mirrors focuses the 960 nm beam in the center of a periodically-poled-potassium-titanyl-phosphate (PPKTP) crystal, while the other one refocusses the 960 nm light while allowing the second harmonic to pass through and exit the cavity. For a maximum input power of 1.4 W, a maximum $400$ mW blue light is obtained experimentally. In order to avoiding instability caused by thermal optical bistability in the SHG cavity, we use a typical power less than 200 mW coupling light for our experiments. The frequency of 960 nm light is locked to the same cavity (the finesse at 960 nm is $\mathcal{F}\approx$20000). Both lasers are locked to a side band which can be used to overlap the corresponding atomic transition frequency with the cavity mode, where the side band is generated by two fiber EOMs working at 780 nm and 960 nm, respectively. The temperature of reference cavity is stabilized to around 32$^\circ$C, and the long drift of the cavity mode is less than 1 MHz/day. Besides, both lasers are set to horizontally polarized inside the vapor cell and their powers are stabilized using a PID controller (based on Redpitaya, STEMlab 125-14) and an acousto-optic modulator (AOM). \\

\textbf{Estimation of the response time of the SR system.}
To estimate the response time of our Rydberg system, we need to characterize the relaxing time of the nonequilibrium dynamics around the critical point. In particular, we sweep the frequency very slow (5 Hz) so that the influence of speed of sweeping frequency is negligible, thus we can measure how fast the system evolves from one of the stable states to another one. Here the results show the evolving time is around 0.5 ms. This result show that when working with equilibrium stable states, the response bandwidth is probably limited by the relaxing time. While working with nonequilibrium states may exhibit other memory effect since the system state is determined not only on the current laser detuning but also previous ones. Here we only consider Markovian stochastic process, and thus we do not adopt signal frequency larger than 2 kHz. On the other hand, the nonequilibrium dynamics may also contribute to deteriorate the synchronization efficiency, resulting in smaller experimental amplification coefficients.\\

\begin{figure}
	\centering
	\includegraphics[scale=0.43]{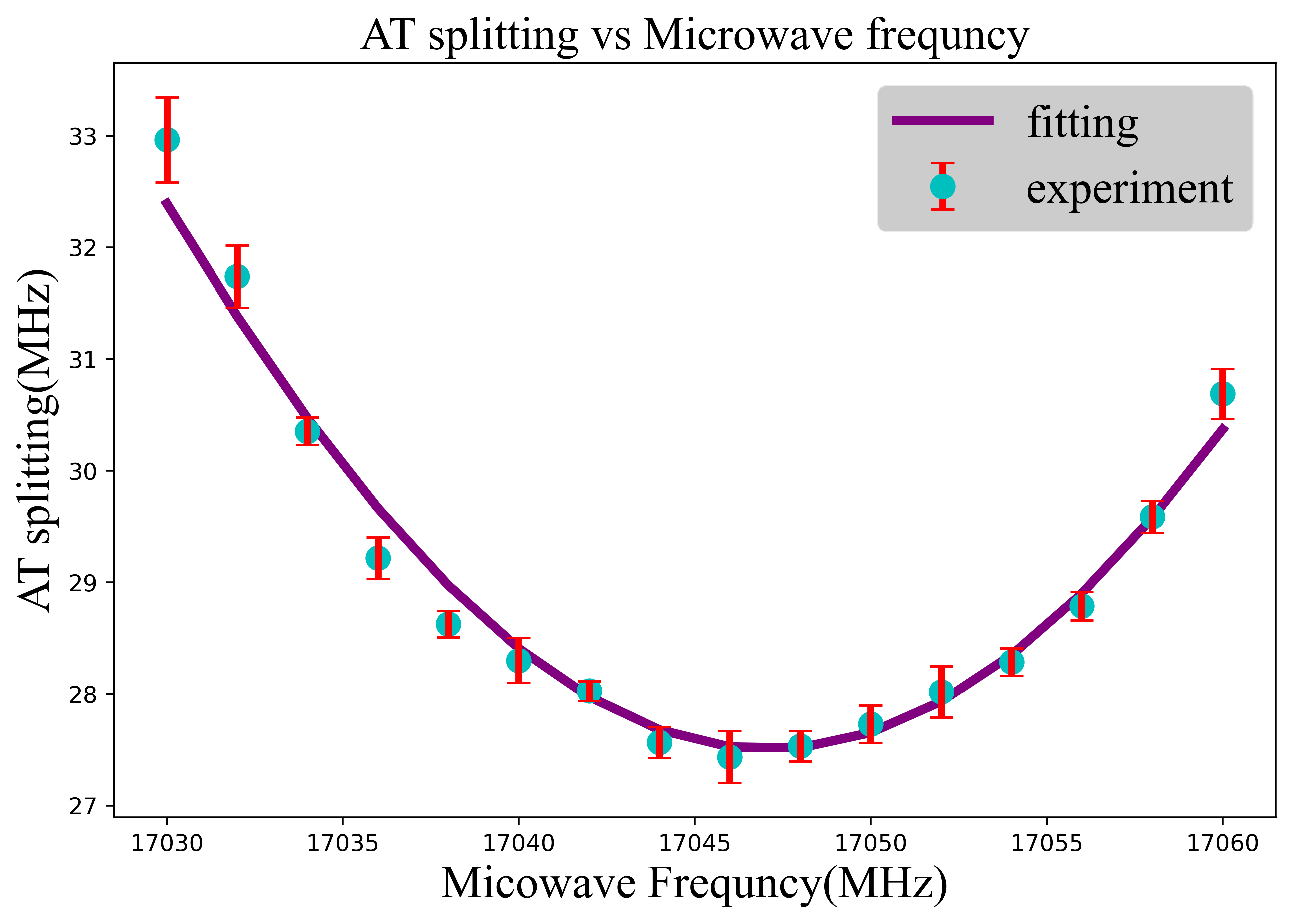}
	\caption{\label{fig:SM1}
		\textbf{Calibration of the MW frequency}. The actual resonant frequency is calibrated, and fit to be 17.047 GHz. The fitting is given by $\Delta_{AT} = \sqrt{27.5^2+(f_{mw}-17047)^2}$.
	}
\end{figure}

\textbf{Calibration of the MW resonant frequency and strength.}
We perform experiments to determine the resonant frequency driving $50D_{5/2}\longleftrightarrow 51P_{3/2}$, and the actual MW strength sensed by the Rydberg atoms. Here we determine the transition frequency in AT splittings regimes by the following steps: (i) Find an approximate transition frequency with the help of ARC\cite{vsibalic2017arc}, for example from 17.03 GHz to 17.06 GHz. (ii) Fix the MW field to a certain strength. (iii) Set the MW to a certain frequency 17.03 GHz, and sweep the coupling laser detunings. (iv) Fit the experimentally obtained PD output with two Gaussian curves, and find center of the two Gaussian curves. (v) Obtain the AT splitting value. (vi) Increase the MW frequency and repeat (iii) to (v). Figure~\ref{fig:SM1} plots the AT splittings against the MW frequency, and we can obtain the resonant frequency by identifying the minimum AT splitting value of the fitted curve $\Delta_{AT} = \sqrt{27.5^2+(f_{mw}-17047)^2}$. The resonant frequency is 17.047 GHz in our experiment. We then determine the relation between MW$_1$ power (in dBm) and actual MW field (in mV/cm) sensed by the atoms using AT splitting method. The results show that $E (\mathrm{mV/cm})=31.09\sqrt{P_{MW}}(\mathrm{mW})$.\\
\begin{figure}
	\centering
	\includegraphics[scale=0.43]{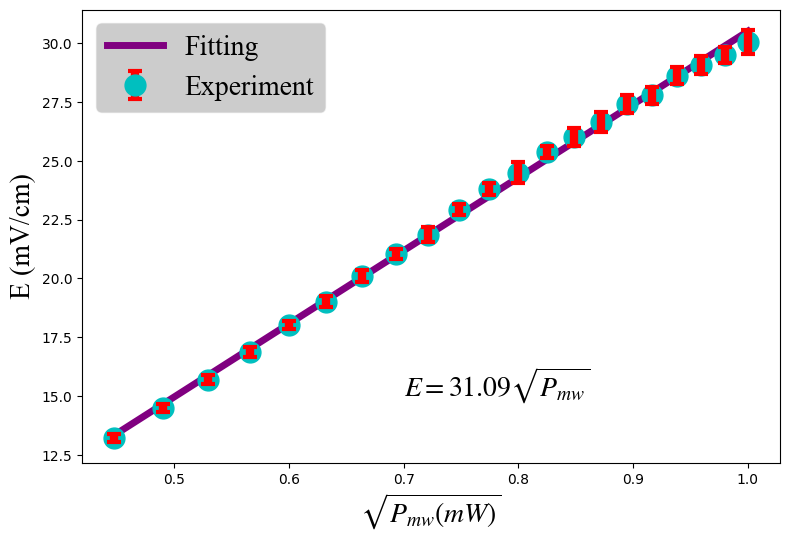}
	\caption{\label{fig:SM2}
		\textbf{Calibration of the MW strength}. The actual strength of the target MW field is calibrated, and fit to be $E (\mathrm{mV/cm})=31.09\sqrt{P_{MW}}(\mathrm{mW})$. 
	}
\end{figure}

\begin{figure}
	\centering
	\includegraphics[scale=0.4]{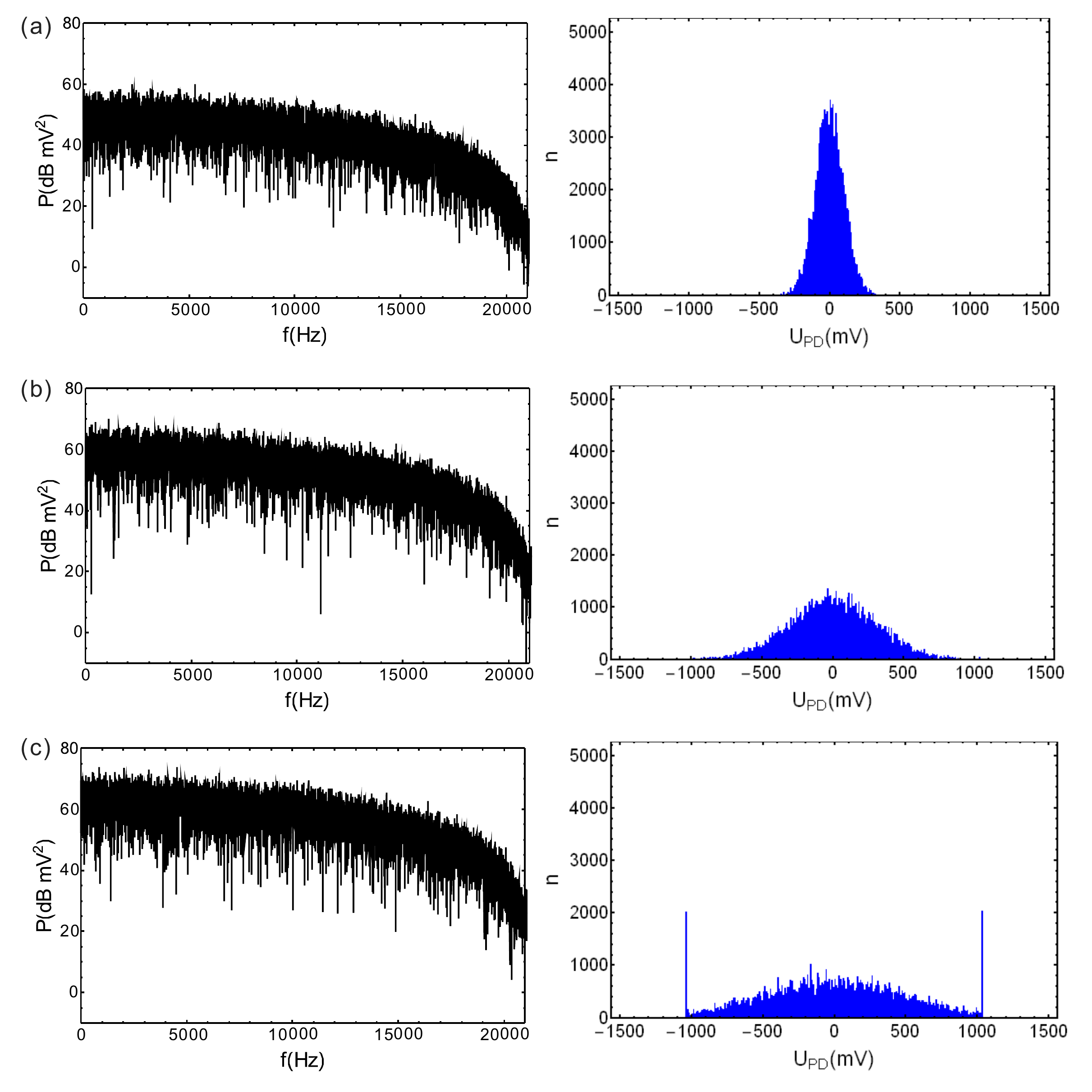}
	\caption{\label{fig:SM3}
		\textbf{White noise spectrum and histogram plot}. (a) D=0.1 V; (b) D=0.3 V, (c) D=0.5 V. 
	}
\end{figure}

\textbf{Generation of the white noise.}
The white noise applied to AM is generated by using Redpitaya (Stem lab 125-14) and the output is continuously generated random voltage ranging from -1V to 1V. Experimentally obtained histogram plot for output voltages are shown in Fig.~\ref{fig:SM3}. Since the output of Redpitaya is limit to $\pm 1$ V, the Gaussian shape will be severely deformed when $D>0.6$V (see Fig.~\ref{fig:SM3}(c) for $D$=0.5). As a result, we set the maximum $D$ to 0.6V (here $D$ is the standard deviation). 

This also put a limitation on the signal frequencies. For lower $f$, smaller $D$ is needed, and thus the experimental results agree with theoretical predictions well. For higher $f$, larger $D$ is required to achieve SR, and thus the performance is degraded, leading to a less efficiency in amplification coefficient $\eta$.  

\end{document}